# DENSITY DISTRIBUTION FOR THE MOLECULES OF A LIQUID IN A SEMI-INFINITE SPACE

VINCENZO MOLINARI

*Alma Mater Studiorum-Università di Bologna*
*Montecuccolino Laboratory of Nuclear Engineering*
*via dei Colli 16, 40136 Bologna (ITALY)*
*vincenzo.molinari@fastwebnet.it*

BARRY D. GANAPOL

*University of Arizona*
*Department of Aerospace and Mechanical Engineering*
*Tucson, Az. 85721 (USA)*
*Ganapol@cowboy.ame.arizona.edu*

DOMIZIANO MOSTACCI[*]

*Alma Mater Studiorum-Università di Bologna*
*Montecuccolino Laboratory of Nuclear Engineering*
*via dei Colli 16, 40136 Bologna (ITALY)*
*domiziano.mostacci@unibo.it*

The Sutherland approximation to the van der Waals forces is applied to the derivation of a self-consistent Vlasov-type field in a liquid filling a half space, bordering vacuum. The ensuing Vlasov equation is then derived, and solved to predict the behavior of the density at and in the vicinity of the liquid-vacuum interface. A numerical solution to the Vlasov equation is also produced and the density profile shown and discussed.

## 1. Introduction

The variation of the number density in a liquid, as a consequence of a boundary discontinuity, is often disregarded, in many applications. Nevertheless, this subject is of growing interest in many fields such as in the study of thin films, e.g. in the petroleum industry or in biology, where investigations are becoming more and more detailed with decreasing dimensions, see, e.g., refs.1-2; also refs.3-6. The aim of our present work is to obtain an estimate of the number density distribution in a liquid as a function of distance from a discontinuity, when the liquid is at rest and in thermodynamic equilibrium, i.e.,

---

[*] corresponding author - e-mail: domiziano.mostacci@unibo.it; tel +39-051-208.77.19; fax: +39-051-208.77.47

the velocity distribution function is Maxwellian with average velocity zero. The setting considered will be a liquid bordering vacuum. The problem of a liquid in contact with its vapour has been treated exhaustively in ref. 7.

The present investigation is conducted in the framework of kinetic theory, aiming to obtain specified liquid properties as brought about by discontinuities. When dealing with a liquid phase, the effects of interaction between molecules can never be disregarded; here an estimate of this effect is obtained resorting to a self-consistent Vlasov field. The self-consistent field is appropriate to describing the dynamic behaviour of a system, where every molecule interacts simultaneously with a large number of surrounding molecules and the correlation between pairs of molecules can be disregarded since the effect of the latter becomes negligible with respect to the collective interaction. This physical situation occurs when the range of the intermolecular force is large in comparison to the average distance between molecules, as is the case of plasma or liquid states. From this point of view, interactions in a liquid system are treated somewhat like is done in plasma physics when there are enough particles in the so-called Debye sphere and simultaneous multiple collisions prevail. To calculate the Vlasov self-consistent field, the intermolecular interaction form is needed: several phenomenological potential models have been proposed to describe the real multiple force interaction mechanism. In this work, the Sutherland pair potential model is used because it is mathematically simple and the predictions on the liquid properties are expected to be qualitatively consistent with the behaviour of many real fluids.

## 2. Governing Equations

The first equation of the of BBGKY hierarchy, describing the microscopic temporal behaviour of the particles of the system in terms of position and velocity, is obtained from the Liouville equation[8] as

$$\frac{\partial f_1}{\partial t} + \vec{v}_1 \cdot \frac{\partial f_1}{\partial \vec{r}_1} + \frac{\vec{F}_1}{m} \cdot \frac{\partial f_1}{\partial \vec{v}_1} = -\int \frac{\vec{F}_{1,2}}{m} \cdot \frac{\partial f_{1,2}}{\partial \vec{v}_1} d\vec{r}_2 d\vec{v}_2 \tag{1}$$

where $f_1(\vec{r}_1,\vec{v}_1,t)$ denotes the single particle (or simple) distribution function, $f_{1,2}(\vec{r}_1,\vec{r}_2,\vec{v}_1,\vec{v}_2,t)$, is the two particle (or double) distribution function, $F_1$ denotes the external force on a particle of mass $m$ located at $(\vec{r}_1,t)$ and $F_{1,2}$ is the inter-particle force between two particles located at positions $\vec{r}_1$ and $\vec{r}_2$. Since the collision integral on the right hand side depends on $f_{1,2}$, the correlation in position and velocity between colliding particles is explicitly considered. This equation is valid up to the order associated with the model used for the intermolecular potential as described below.

Consider a system in steady thermodynamic equilibrium and whose particles are not subject to an external force, so that $\partial f_1/\partial t = 0$ and $F_1 = 0$. In addition, assume that the inter-particle force $F_{1,2}$ is given by a central potential independent of velocity. In this case, assuming molecular chaos, the double distribution function, $f_{1,2}$, is given by

$$f_{1,2}(\vec{v}_1,\vec{v}_2) = n_{1,2} f_M(v_1) f_M(\vec{v}_2), \tag{2}$$

where $f_M(\vec{v})$ denotes a Maxwellian distribution and $n_{1,2}(\vec{r}_1, \vec{r}_2)$ denotes the two-particle number density

$$n_{1,2}(\vec{r}_1, \vec{r}_2) = \int f_{1,2} d\vec{v}_1 d\vec{v}_2 \ . \tag{3}$$

Observing that

$$\frac{\partial f_M}{\partial \vec{v}_1} = -\frac{m}{K_B T} \vec{v}_1 f_M(\vec{v}_1), \tag{4}$$

where $K_B$ is the Boltzmann constant and $T$ is the (uniform) temperature, Eq. (1) becomes

$$K_B T \frac{\partial n_1}{\partial \vec{r}_1} = \int n_{1,2} \vec{F}_{1,2} d\vec{r}_2 \ , \tag{5}$$

where $n_1(\vec{r}_1)$ is the single particle density

$$n_1(\vec{r}_1) = \int n_{1,2} d\vec{r}_2 \ . \tag{6}$$

When correlation between pairs of particles can be disregarded, the two-particle number density simplifies significantly:

$$n_{1,2} \approx n_1 n_2 \ . \tag{7}$$

The self-consistent field is important to describe the dynamic behaviour of a system, where every molecule interacts simultaneously with a large number of surrounding molecules and the correlation between pairs of molecules can be disregarded since its effect becomes negligible compared to the collective interaction. This physical situation occurs when the range of the intermolecular force, $F_{1,2}$, is large in comparison to the average distance between molecules, as is the case of a plasma or liquid state. From this point of view, interactions in a liquid system are treated somewhat like is done in plasma physics, where a specific parameter, the "plasma parameter", establishes when there are enough particles in the so-called "Debye sphere" and simultaneous multiple collisions prevail. Then, for the liquid state, we can assume Eq. (7) to be valid and the collision integral becomes

$$\int n_{1,2} \vec{F}_{1,2} d\vec{r}_2 = n_1 \int n_2 \vec{F}_{1,2} d\vec{r}_2 = n_1 \vec{F}' \ , \tag{8}$$

where $\vec{F}'$ is the unknown self-consistent field that must be determined on the basis of the pairwise interaction potential. With this potential, Eq. (5) can be rewritten as

$$K_B T \frac{1}{n_1} \frac{\partial n_1}{\partial \vec{r}_1} = \vec{F}'(\vec{r}_1) \tag{9}$$

.

V Molinari, BD Ganapol, D Mostacci

## 2.1. *Self-consistent field*

The next step is to find the self-consistent field, $\vec{F}'$ from

$$\vec{F}'(\vec{r}_1) = \int_V n(\vec{r}_2) \frac{\partial \varphi_{1,2}}{\partial \vec{r}_1} dr_2, \qquad (10)$$

where $\varphi_{1,2}$ is the pairwise interaction potential between two molecules located at positions $\vec{r}_1$ and $\vec{r}_2$ respectively. To calculate the Vlasov self-consistent field, several phenomenological potential models have been proposed which, more or less, describe the real multiple force interaction mechanism[9-10]. In this work, the Sutherland pair potential model is used for its mathematical simplicity and because the predictions of the liquid properties are expected to be qualitatively consistent with the behaviour of many real fluids.

The phenomenological Sutherland pair potential model given by[9]

$$\begin{cases} \varphi_{1,2}(r) = -4\varepsilon \left(\frac{\sigma}{r}\right)^\alpha & \text{for } r \geq \sigma \\ \varphi_{1,2}(r) = \infty & \text{for } r < \sigma \end{cases} \qquad (11)$$

with the corresponding force $F_{1,2}$

$$\vec{F}_{1,2} = -\vec{\nabla}\varphi_{1,2} = 4\varepsilon\alpha \frac{\sigma^\alpha}{r^{\alpha+1}} \hat{r}. \qquad (12)$$

The Sutherland parameters $\varepsilon$ and $\sigma$ correspond, respectively, to the depth of the potential well (divided by 4) and to the distance of closest approach in the Sutherland approximation, essentially related to the molecular radius. Both parameters are tabulated for numerous molecules, see e.g. ref. 9.

To calculate $\vec{F}'$, a molecule located at the point $(0,0,z)$ will be considered, and the force from all the surrounding liquid will be determined using the Sutherland potential, assuming slab symmetry, i.e., density depends only on the z – coordinate, and considering an elementary volume $dV$ at a location defined by a spherical coordinate system centered on the molecule of interest. By expressing the Cartesian force components and integrating the forces over the all space, after some algebra, the self-consistent field for the semi-infinite medium is found to be[11]

$$F' = 8\pi\varepsilon\sigma^6 \left\{ -\int_0^{z-\sigma} \frac{n(\zeta)}{(z-\zeta)^5} d\zeta + \int_{z+\sigma}^{+\infty} \frac{n(\zeta)}{(\zeta-z)^5} d\zeta \right\}. \qquad (13a)$$

For completeness, the same procedure applied to a slab of thickness 2a would yield

$$F' = 8\pi\varepsilon\sigma^6 \left\{ -\int_{-a}^{z-\sigma} \frac{n(\zeta)}{(z-\zeta)^5} d\zeta + \int_{z+\sigma}^{a} \frac{n(\zeta)}{(\zeta-z)^5} d\zeta \right\}. \qquad (13b)$$

If the density variation is mild, $n(z)$ can be expanded in a Taylor series, and retaining only the first few terms gives

$$n(\zeta) = n(z) + \frac{dn(z)}{dz}(\zeta - z) + \frac{d^2n(z)}{dz^2}\frac{(\zeta-z)^2}{2} + \frac{d^3n(z)}{dz^3}\frac{(\zeta-z)^3}{3!} + \frac{d^3n(z)}{dz^3}\frac{(\zeta-z)^4}{4!} + O\left[(\zeta-z)^5\right]. \quad (14)$$

Neglecting terms of order 3 and higher, and substituting into Eqs. (13) (after some more algebra), the following equations are obtained:

$$F'(z) = 8\pi\varepsilon\sigma^6\left\{\frac{n}{4z^4} + \frac{dn}{dz}\frac{1}{3}\left(\frac{2}{\sigma^3} - \frac{1}{z^3}\right) + \frac{d^2n}{dz^2}\frac{1}{4z^2}\right\} \quad (15a)$$

for the semi-infinite case and, again for completeness,

$$F'(z) = 8\pi\varepsilon\sigma^6\left\{\frac{n}{4(z+a)^4} - \frac{1}{4(a-z)^4} + \frac{dn}{dz}\frac{1}{3}\left(\frac{2}{\sigma^3} - \frac{1}{(z+a)^3} - \frac{1}{(a-z)^3}\right) + \right.$$
$$\left. + \frac{d^2n}{dz^2}\frac{1}{4}\left(\frac{1}{(z+a)^2} - \frac{1}{(a-z)^2}\right)\right\}, \quad (15b)$$

for a slab of thickness 2a.

## 3. The Semi-Infinite Liquid

Consider a liquid occupying a semi-infinite space, $z \geq 0$. The other part of the space, $z < 0$, can be vacuum, or a material in gaseous, liquid or solid phase. Only the first case will be considered here: z < 0 is vacuum.

The equation for the number density distribution, Eq. (9), for the one dimensional case - and dropping from now on the subscript 1 - becomes

$$\frac{dn}{dz} = \frac{n}{KT}F'(n) \quad (16)$$

and substituting the expression of $F'$ in Eq. (16) gives, after some algebra,

$$nn'' + n'z^2\left[n\frac{4}{3}\left(\frac{2}{\sigma^3} - \frac{1}{z^3}\right) - \frac{K_BT}{2\pi\varepsilon\sigma^6}\right] + \frac{n^2}{z^2} = 0 \quad (17a)$$

that is, a nonlinear, singular differential equation of the second order. This equation can be recast into a self-similar form defining the following dimensionless quantities $x = z/\sigma$ ; $\chi = n/n_0$, where $n_0 = n(z = \infty)$ to give

$$\chi\chi'' + \chi'x^2\left[\chi\left(\frac{8}{3} - \frac{4}{3x^3}\right) - \frac{K_BT}{2\pi\varepsilon n_0\sigma^3}\right] + \frac{\chi^2}{x^2} = 0. \quad (17b)$$

For the usual values of $\sigma$ and $\varepsilon$ the quantity $\dfrac{K_B T}{2\pi\varepsilon n_0 \sigma^3}$ can be disregarded (at 20 °C e.g., for water it is 0.094, for ethanol 0.13, for acetone 0.13) and Eq. (17b) becomes

$$x^2 \frac{d^2\chi}{dx^2} + x\left(\frac{8}{3}x^3 - \frac{4}{3}\right)\frac{d\chi}{dx} + \chi = 0 \qquad (18)$$

With the additional variable transformations: $\xi = x^3$, then $\chi = \xi^3 \psi(\xi)$, the confluent hypergeometric equation is obtained, which has a known solution. Reverting to the original variables, the solution to Eq. (18) is found to be

$$\chi(x) = C_1 x^{3k} e^{-Bx} \Phi(A_1 - k, A_1; B_0 x^3) \qquad (19)$$

where $\Phi$ is the confluent hypergeometric function[12] and $k$ is the larger solution of

$$k^2 + (A_0 - 1)k + a = 0 \qquad (20)$$

with

$$A_0 = \frac{2}{9}, \; B_0 = \frac{8}{9}, \; a = \frac{1}{9}, \; A_1 = A_0 + 2k . \qquad (21)$$

To find $C_1$, the following asymptotic behaviour is used (Abramowitz, 1964)

$$\chi(x) = C_1 \frac{\Gamma(A_1)}{B_0^k \Gamma(A_1 - k)} \left\{ 1 + \frac{(1 - A_1 + k)k}{B_0 x^3} + \cdots \right\} \qquad (22)$$

and since $\chi(x \to \infty) = 1$, one finds

$$C_1 = \frac{\Gamma(A_1 - k) B_0^k}{\Gamma(A_1)} . \qquad (23)$$

## 4. Numerical Solution

The numerical scheme begins with Eq. (18) in the more convenient form

$$t^2 \frac{d^2\psi(t)}{dt^2} + t(A_0 + B_0 t)\frac{d\psi(t)}{dt} + a\psi(t) = 0 \qquad (24)$$

with substitutions $t = x^3$, $\chi(x) = \psi(t)$ and $A_0 = 2/9$, $B_0 = 8/9$ and $a = 1/9$.

Since the ODE is singular, special attention must be given to the singularity at the origin, which comes about because of the discontinuity in material properties. Once the singular nature of Eq. (24) has been accommodated through the leading non-analytic factor, for $0 < t$, the solution is expected to be infinitely smooth and therefore a Taylor series should hold. Hence, we apply the method of continuous analytical continuation (CAC) to provide a highly accurate solution for $t$ greater than zero.

Following the method of Frobenius near the origin, the algorithm begins with the solution

$$\psi(t) = t^k f(t), \tag{25a}$$

where

$$f(t) = \sum_{n=0}^{\infty} \psi_n t^n . \tag{25b}$$

By substitution, the coefficients can be shown to be

$$\psi_n = \left[ \frac{B_0(n + k_\pm - 1)}{(n + k_\pm)(n + k_\pm - 1) + A_0(n + k_\pm) + a} \right] \psi_{n-1} \tag{26a}$$

with

$$k_\pm = \frac{1}{2}\left[ -(A_0 - 1) \pm \sqrt{(A_0 - 1)^2 - 4a} \right] . \tag{26b}$$

$\psi_n$ is taken as unity since the solution is to be normalized to unity for large $t$. Since the discontinuity is only in the first interval, the solution is expected to follow the Taylor series,

$$\psi(t) = \sum_{n=0}^{\infty} \psi_{n,j}(t - t_{j-1})^n , \tag{27}$$

thereafter in any interval $t_{j-1} \leq t \leq t_j$. The coefficients $\psi_{n,j}$ are found by substitution of Eq. (27) into Eq. (24) to give (after some algebra) for $n = 1,2$

$$\begin{aligned}
2t_{j-1}^2 \psi_{2,j} &= -t_{j-1}(A_0 + B_0 t_{j-1})\psi_{1,j} - a\psi_{0,j} \\
6t_{j-1}^2 \psi_{3,j} &= (A_0 + a + 2B_0 t_{j-1})\psi_{1,j} - [4t_{j-1} + 2t_{j-1}(A_0 + B_0 t_{j-1})]\psi_{2,j}
\end{aligned} \tag{28a}$$

and for $n \geq 4$

$$\begin{aligned}
n(n-1)t_{j-1}^2 \psi_{n,j} &= -[(n-2)(n-3) + (n-2)A_0 + a + 2B_0 t_{j-1}(n-2)]\psi_{n-2,j} - \\
&\quad -[2t_{j-1}(n-1)(n-2) + t_{j-1}(A_0 + B_0 t_{j-1})(n-1)]\psi_{n-1,j}
\end{aligned} \tag{28b}$$

Finally, the function and its derivative at the end of the interval give the initial two coefficients

$$\psi_{0,j+1} = \psi(t_j) = \sum_{n=0}^{\infty} \psi_{n,j}(t_j - t_{j-1})^n \tag{29c}$$

$$\psi_{1,j+1} = \frac{d\psi(t)}{dt}\bigg]_{t=t_j} = \sum_{n=1}^{\infty} n\psi_{n,j}(t_j - t_{j-1})^{n-1} \tag{29d}$$

which connect to the next interval completing one entire interval calculation. The recurrence is initiated from the first interval with

$$\psi_{0,2} = \psi_0(t_1) = t_1^{k_+} \sum_{n=0}^{\infty} \psi_{n,1} t_1^n$$

$$\psi_{1,2} = \psi_1(t_1) = \frac{k_+}{t_1} \psi_0 + t_1^{k_+} \sum_{n=1}^{\infty} n \psi_{n,1} t_1^{n-1}$$

(29d)

To provide confidence in the numerical procedure, Table 1 gives the comparison for large values of $t$ for the un-normalized solution. As observed, with increasing $t$, the asymptotic form indeed emerges. This calculation required that each interval be further divided into 4 converged sub-intervals for 9-place precision.

Table 1. Comparison to Asymptotic Solution

| t | $\psi(t_j)$ | $\psi_{asy}(t_j)$ | Identical Digits |
|---|---|---|---|
| 5.0000E+02 | 1.87080215E+00 | 1.87078599E+00 | 5 |
| 1.0000E+03 | 1.86985554E+00 | 1.86985153E+00 | 5 |
| 1.5000E+03 | 1.86954182E+00 | 1.86954005E+00 | 6 |
| 2.0000E+03 | 1.86938530E+00 | 1.86938430E+00 | 6 |
| 2.5000E+03 | 1.86929149E+00 | 1.86929086E+00 | 6 |
| 3.0000E+03 | 1.86922900E+00 | 1.86922856E+00 | 6 |
| 4.0000E+03 | 1.86915094E+00 | 1.86915069E+00 | 6 |
| 5.0000E+03 | 1.86910412E+00 | 1.86910396E+00 | 7 |
| 6.0000E+03 | 1.86907293E+00 | 1.86907282E+00 | 7 |

## 5. Conclusions

First, recall that the case considered is that of an interface between a liquid and vacuum. Molecules in the bulk of the homogeneous liquid are subject to binding forces having no preferential direction, or in other words, isotropic forces. Hence density is to be expected constant in regions far enough from interfaces. However, as an interface is approached (i.e., at distances of the order of the tens of molecular diameters) binding forces become peaked toward the bulk of the liquid, and it can be expected that this prevailing inward force would lead to an increase in the density, which is the opposite of what is commonly assumed. The aim of the present work was to verify the existence of this phenomenon: Figure 1 confirms this prediction, insofar as the attractive forces present are of the van der Waals type. This behaviour will be of particular importance in investigating surface tension.

If an interface with a solid surface is considered, on the other hand, this behaviour can be significantly changed, or even reversed, depending upon the relative forces between the molecules belonging to the two phases. This is the subject of further work already in progress.

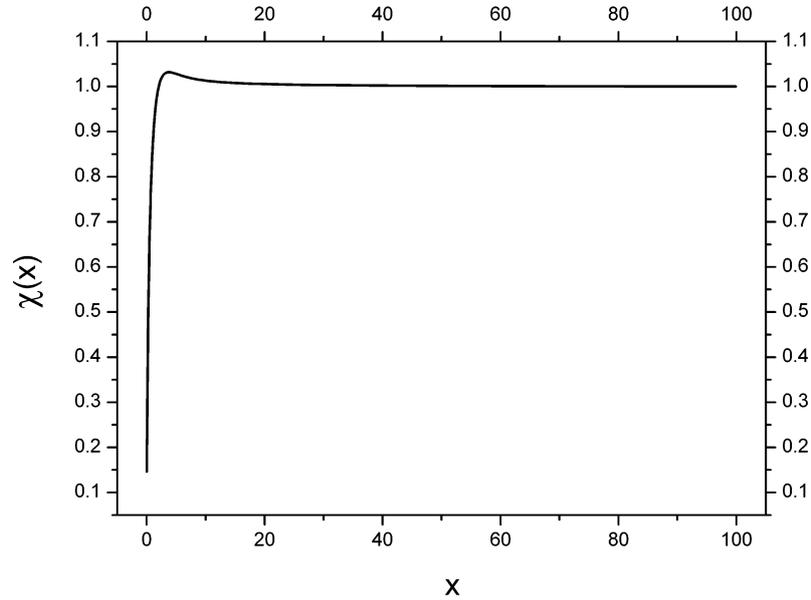

Figure 1. Dimensionless density vs. dimensionless distance from the interface.

## Acknowledgements

The second author thanks the University of Bologna Institute for Advanced Study for hosting his stay during the completion of this work